\documentclass{emulateapj}
\def\degpoint{\ifmmode ^{\rm{o}}\!. \else $^{\rm{o}}\!.$\fi}

\newcommand{\ms}{\mbox{m\ s$^{-1}$}}

\newcommand{\Mjup}{\mbox{M$_{\rm Jup}$}}

\newcommand{\ltsimeq}{\raisebox{-0.6ex}{$\,\stackrel
         {\raisebox{-.2ex}{$\textstyle <$}}{\sim}\,$}}
\newcommand{\gtsimeq}{\raisebox{-0.6ex}{$\,\stackrel
         {\raisebox{-.2ex}{$\textstyle >$}}{\sim}\,$}}

\begin{document}
\title{Dynamical and Observational Constraints on Additional Planets in 
Highly Eccentric Planetary Systems\footnotemark[1]}
\author{Robert A.~Wittenmyer, Michael Endl, William D.~Cochran}
\affil{McDonald Observatory, University of Texas at Austin, Austin, TX 
78712}
\email{
robw@astro.as.utexas.edu }
\author{Harold F.~Levison}
\affil{Department of Space Studies, Southwest Research Institute, Boulder, 
CO 80302}

\shorttitle{Highly Eccentric Planetary Systems}
\shortauthors{Wittenmyer et al.}
\slugcomment{Accepted for publication in AJ}
\footnotetext[1]{Based on observations obtained with the Hobby-Eberly
Telescope, which is a joint project of the University of Texas at Austin,
the Pennsylvania State University, Stanford University,
Ludwig-Maximilians-Universit{\" a}t M{\" u}nchen, and
Georg-August-Universit{\" a}t G{\" o}ttingen.}
\begin{abstract}

\noindent Long time coverage and high radial velocity precision have
allowed for the discovery of additional objects in known planetary
systems.  Many of the extrasolar planets detected have highly eccentric
orbits, which raises the question of how likely those systems are to host
additional planets.  We investigate six systems which contain a very
eccentric ($e>0.6$) planet: HD~3651, HD~37605, HD~45350, HD~80606,
HD~89744, and 16~Cyg~B.  We present updated radial-velocity observations
and orbital solutions, search for additional planets, and perform test
particle simulations to find regions of dynamical stability.  The
dynamical simulations show that short-period planets could exist in the
HD~45350 and 16~Cyg~B systems, and we use the observational data to set
tight detection limits, which rule out additional planets down to a few
Neptune masses in the HD~3651, HD~45350, and 16~Cyg~B systems.

\end{abstract}


\section{Introduction}

One surprising result that has come out of the more than 200 extrasolar
planet discoveries to date is the wide range of eccentricities observed.  
Unlike our own Solar system, many of the extrasolar planets which are not
tidally locked to their host stars have moderate eccentricities ($e>0.2$),
and 15 planets have high eccentricities ($e>0.6$).  These observations
have spawned several theories as to the origin of highly eccentric
extrasolar planets.  One such method, planet-planet scattering, occurs
when multiple jovian planets form several astronomical units (AU) from the
host star and then interact, leaving one in an eccentric orbit and often
ejecting the other \citep{rasio96}.  This method has been proposed to
explain the architecture of the $\upsilon$ And planetary system
\citep{ford05}, which contains a hot Jupiter as well as two jovian planets
in moderately eccentric orbits.  \citet{linida97} suggested a merger
scenario in which inner protoplanets perturb each other and merge to form
a single massive, eccentric planet with $e\gtsimeq 0.3$ and $a\sim 0.5-1$
AU.

Interactions with stellar companions are another possible way to boost a
planet's eccentricity.  Of the 15 stars hosting a planet with $e>0.6$, six
are also known to possess stellar-mass companions in wide binary orbits:
HD~3651 \citep{mugrauer06, luhman07}, HD~20782 \citep{desidera07},
HD~80606, HD~89744 \citep{wilson01, mugrauer04}, 16~Cyg~B, and HD~222582
\citep{raghavan06}.  If the inclination angle between the planetary orbit
and a stellar companion is large, the Kozai mechanism \citep{kozai} can
induce large-amplitude oscillations in the eccentricity of the planet
(e.g.~Malmberg et al.~2006).  These oscillations can be damped by general
relativistic effects and by interaction with other planets, and hence are
most effective in systems with a single planet in an orbit $a\gtsimeq$1~AU
from the host star \citep{takeda05}.  The Kozai mechanism has been
suggested to explain the high eccentricity of 16~Cyg~Bb \citep{holman97,
mazeh97} and HD~80606b \citep{wu03}.  \citet{hauser99} found the
inclination of 16~Cyg~B orbiting the system barycenter to lie between 100
and 160 degrees, where 90 degrees is an edge-on orientation.  However, it
is the difference in inclination between the orbital planes of the
planetary and stellar companion that is critical in determining the
importance of the Kozai mechanism, and the inclination of the planet's
orbit is generally not known for non-transiting systems.

Of the 192 known planetary systems, 23 (12\%) are multi-planet systems.  
Recent discoveries of additional objects in systems known to host at least
one planet \citep{udry07, destructor, rivera05, vogt05, mcarthur04,
santos04} suggest that multiple-planet systems are common.  Of particular
interest are systems which host a jovian planet and a low-mass ``hot
Neptune,'' e.g.~55~Cnc (=HD~75732), GJ~876, $\mu$ Arae (=HD~160691),
Gl~777A (=HD~190360).  Motivated by the discoveries of hot Neptunes in
known planetary systems, we have undertaken an intensive survey of
selected single-planet systems to search for additional low-mass
companions.  Three of the planetary systems discussed in this paper
(HD~3651, HD~80606, HD~89744) are part of this campaign.  The excellent
radial-velocity precision of the High Resolution Spectrograph on the
Hobby-Eberly Telescope (HET), combined with queue-scheduling, allow us to
time the observations in such a way as to minimize phase gaps in the orbit
of the known planet, and also to act quickly on potential new planet
candidates.  The use of the HET in this manner is discussed further in
\citet{cochran04} with regard to the discovery of HD~37605b.

In this work, we aim to combine observational limits on additional planets
in known planetary systems with dynamical constraints obtained by N-body
simulations.  The observations address the question: What additional
planets are (or are not) in these systems?  The dynamical simulations can
answer the question: Where are additional planets possible?  Section~2
describes the observations and the test particle simulations for six
highly eccentric planetary systems: HD~3651, HD~37605, HD~45350, HD~80606,
HD~89744, and 16~Cyg~B.  We have chosen these systems based on two
criteria: (1) Each hosts a planet with $e>0.6$, and (2) Each has been
observed by the planet search programs at McDonald Observatory.  In \S 3,
we present and discuss the results of the updated orbital fits, dynamical
simulations, and detection limit computations.

\section{Observations and Data Analysis}

\subsection{Radial-Velocity Observations}

Five of the six stars considered in this work have been observed with the
McDonald Observatory 9.2~m Hobby-Eberly Telescope (HET) using its High
Resolution Spectrograph (HRS) \citep{tull98}.  A full description of the
HET planet search program is given in \citet{cochran04}.  For 16~Cyg~B,
observations from McDonald Observatory were obtained only with the 2.7~m
Harlan J.~Smith (HJS) telescope; the long-term planet search program on
this telescope is described in \citet{limitspaper}.  All available
published data on these systems were combined with our data from McDonald
Observatory in the orbit fitting procedures.

\subsection{Numerical Methods}

To place constraints on the architecture of planetary systems, we would
like to know where additional objects can remain in stable orbits in the
presence of the known planet(s).  We performed test particle simulations
using SWIFT\footnote{SWIFT is publicly available at
http://www.boulder.swri.edu/$\sim$hal/swift.html.} \citep{levison94} to
investigate the dynamical possibility of additional low-mass planets in
each of the six systems considered here.  Low-mass planets can be treated
as test particles since the exchange of angular momentum with jovian
planets is small.  We chose the regularized mixed-variable symplectic
integrator (RMVS3) version of SWIFT for its ability to handle close
approaches between massless, non-interacting test particles and planets.  
Particles are removed if they are (1) closer than 1 Hill radius to the
planet, (2) closer than 0.05~AU to the star, or (3)  farther than 10~AU
from the star.  Since the purpose of these simulations is to determine the
regions in which additional planets could remain in stable orbits, we set
this outer boundary because the current repository of radial-velocity data
cannot detect objects at such distances.

The test particle simulations were set up following the methods used in
\citet{barnes04}, with the exception that only initially circular orbits
are considered in this work. For each planetary system, test particles
were placed in initially circular orbits spaced every 0.002~AU in the
region between 0.05-2.0~AU.  We have chosen to focus on this region
because the duration of our high-precision HET data is currently only 2-4
years for the objects in this study.  The test particles were coplanar
with the existing planet, which had the effect of confining the simulation
to two dimensions.  Input physical parameters for the known planet in each
system were obtained from our Keplerian orbit fits described in \S~3.1,
and from recent literature for 16~Cyg~B \citep{destructor} and HD~45350
\citep{endl06}.  The planetary masses were taken to be their minimum
values (sin~$i=1$).  The systems were integrated for $10^7$ yr, following
\citet{barnes04} and allowing completion of the computations in a
reasonable time.  We observed that nearly all of the test-particle
removals occurred within the first $10^6$ yr; after this time, the
simulations had essentially stabilized to their final configurations.

\section{Results and Discussion}

\subsection{Updated Keplerian Solutions for 4 Systems}

We present updated Keplerian orbital solutions for HD~3651b, HD~37605b,
HD~80606b, and HD~89744b in Table~1.  A summary of the data used in our
analysis is given in Table~2, and the HET radial velocities are given in
Tables~3-6.  The velocity uncertainties given for the HET data represent
internal errors only, and do not include any external sources of error
such as stellar ``jitter.'' The parameters for the remaining two planets,
HD~45350b and 16~Cyg~Bb, are taken from \citet{endl06} and
\citet{destructor}, respectively.  Radial velocity measurements from the
HET are given for HD~45350 in \citet{endl06}, and velocities for 16~Cyg~B
from the HJS telescope are given in \citet{destructor}.  As in
\citet{destructor}, all available published data were combined with those
from McDonald, and the known planet in each system was fit with a
Keplerian orbit using GaussFit \citep{jefferys87}, allowing the velocity
offset between each data set to be a free parameter.  Examination of the
residuals to our Keplerian orbit fits revealed no evidence for additional
objects in any of the six systems in this study.

The Saturn-mass (M~sin~$i=0.2$\Mjup) planet HD~3651b was discovered by
\citet{fischer03} using observations from Lick and Keck.  We fit these
data, which were updated in \citet{butler06}, in combination with
observations from the HJS and HET at McDonald Observatory. The HET data,
which consist of multiple exposures per visit, were binned using the
inverse-variance weighted mean value of the velocities in each visit.  
The standard error of the mean was added in quadrature to the weighted rms
about the mean velocity to generate the error bar of each binned point
(N=29).  The rms about the combined fit for each dataset is: Lick \&
Keck--6.6 \ms, HET--9.4 \ms, HJS--12.2 \ms.  The fitted orbital parameters
for HD~3651b are of comparable precision to those reported in
\citet{butler06}, and agree within 2$\sigma$.  The recent discovery of a T
dwarf companion to HD~3651 \citep{mugrauer06, luhman07} prompts an
interesting exercise: Can the radial-velocity trend due to this object be
detected in the residuals after removing the planet?  We detect a slope of
$-0.27\pm0.05$ \ms~yr$^{-1}$, indicating that we are indeed able to
discern a trend which is possibly due to the binary companion.  However,
the reduced $\chi^{2}$ of the orbital solution is not significantly
improved by the inclusion of a linear trend ($\Delta
\chi^{2}_{\nu}$=0.18).  The parameters given in Table~1 were obtained from
the fit which did not include a trend.

We present 23 new HET observations for HD~37605 obtained since its
announcement by \citet{cochran04}.  The data now span a total of
1065~days.  The best fit is obtained by including an acceleration of
$-20.5\pm 2.1$ \ms~yr$^{-1}$, indicating a distant orbiting body.  Such a
finding would lend support to the hypothesis that very eccentric
single-planet systems originate by interactions within a wide binary
system.  The shortest period that this outer companion could have and
still remain consistent with the observed acceleration and its uncertainty
over the timespan of the observations is about 40~yr, assuming a circular
orbit.  This object would then have a minimum mass in the brown dwarf
range.

The planet orbiting HD~80606, first announced by \citet{naef01}, is the
most eccentric extrasolar planet known, with $e=0.933\pm 0.001$ (Table~1).  
We have fit the CORALIE data in combination with the Keck data given in
\citet{butler06} and 23 observations from HET.  The extreme velocity
variations induced by this planet greatly increase the sensitivity of
orbit fits to the weighting of individual measurements.  Since the
uncertainties of the HET velocities given in Tables~3-6 represent internal
errors only, we experimented with adding 1-7 \ms of radial-velocity
``jitter'' in quadrature before fitting the data for HD~80606.  For all of
these jitter values, the fitted parameters remained the same within their
uncertainties.  Table~1 gives the parameters derived from a fit which
added 3.5~\ms of jitter \citep{butler06} to the HET data.  The rms about
the combined fit is: CORALIE--18.7 \ms, HET--7.5 \ms, Keck--5.6 \ms.  
\citet{butler06} noted that the eccentricity $e$ and the argument of
periastron $\omega$ had to be held fixed in their fit to the Keck data
alone.  However, the large number of measurements included in this work
allowed GaussFit to converge with all parameters free.

For HD~89744b, we combine data from the HET with 6 measurements from the
HJS telescope and Lick data from \citet{butler06}.  The HET data were
binned in the same manner as for HD~3651, resulting in N=33 independent
visits.  The rms about the combined fit for each dataset is: Lick--17.1
\ms, HET--10.7 \ms, HJS--9.5 \ms.  As with HD~3651b, our derived
parameters agree with those of \citet{butler06} within 2$\sigma$.  The
scatter about our fit remains large, most likely due to the star's early
spectral type (F7V), which hinders precision radial-velocity measurements
due to the smaller number of spectral lines.  For example, the F7V star
HD~221287 was recently found to host a planet \citep{naef07};  despite the
superb instrumental precision of the HARPS spectrograph, that orbital
solution has a residual rms of 8.5 \ms.

\subsection{Test Particle Simulations}

The results of the dynamical simulations are shown in Figures~1-3.  The
survival time of the test particles is plotted against their initial
semimajor axis.  As shown in Figure~1, the short-period planets HD~3651
and HD~37605 sweep clean the region inside of about 0.5~AU.  In both of
these systems, however, a small number of test particles remained in
low-eccentricity orbits near the known planet's apastron distance, near
the 1:2 mean-motion resonance (MMR).  In the HD~3651 system, particles
remained stable beyond about 0.6~AU, which is not surprising given the low
mass of the planet.  For HD~37605, two distinct strips of stability are
seen in Fig.~1, corresponding to the 1:2 and 1:3 MMRs.  The eccentricity
of the test particles in the region of the 1:2 MMR oscillated between 0.00
and 0.06.  Particles in 1:3 MMR oscillated in eccentricity with a larger
range, up to $e\sim 0.4$, which is expected due to secular forcing. As
with HD~3651, the region beyond about 0.8~AU was essentially unaffected by
the planet.

Figure~2 shows the results for the HD~45350 and HD~80606 systems.  The
long period (963.6 days) and relatively large mass (M~sin~$i$=1.8 \Mjup)  
of HD~45350b restricted stable orbits to the innermost 0.2~AU.  These test
particles oscillated in eccentricity up to $e\sim 0.25$.  The
4\Mjup~planet orbiting HD~80606 removed all test particles to a distance
of about 1.5~AU, and only beyond 1.75~AU did test particles remain in
stable orbits for the duration of the simulation ($10^7$ yr).  A region of
instability is evident at 1.9~AU due to the 8:1 MMR.  Figure~3 shows that
HD~89744b eliminated all test particles except for a narrow region near
the 8:3 resonance.  For the 16~Cyg~B system, only particles inside of
about 0.3~AU remained stable, leaving open the possibility of short-period
planets.  The surviving particles oscillated in eccentricity up to $e\sim
0.45$, but these simulations treat the star as a point mass, and hence
tidal damping of the eccentricity is not included.  Our results are
consistent with those of \citet{menou03}, who investigated dynamical
stability in extrasolar planetary systems and found that no test particles
survived in the habitable zones of the HD~80606, HD~89744, and 16~Cyg~B
systems.

\subsection{Detection Limits}

Three of these systems (HD~3651, HD~80606, HD~89744) were monitored
intensely with the HET as part of a larger effort to search for low-mass,
short period planets.  No evidence was found for any such objects in these
or any of the six systems in this work.  We then asked what limits can be
set on additional planets using the high-precision HET data we have
obtained.  The procedure for determining companion limits was identical to
the method described in \citet{limitspaper}, except that in this work, the
best-fit Keplerian orbit for the known planet (see \S~3.1) was removed
before performing the limits computations.  In this way, we determined the
radial-velocity amplitude $K$ for which 99\% of planets would have been
detected in the residuals.  The eccentricity of the injected test signals
was chosen to be the mean eccentricity of the surviving particles from the
simulations described in \S~3.2.  Only the regions in which test particles
survived were considered in these limits computations.

The results of these computations were highly varied, reflecting the
differing observing strategies employed for these six objects.  In
particular, HD~3651, HD~80606, and HD~89744 were monitored intensely with
the HET as part of a search for short-period objects, whereas HD~37605 and
HD~45350 were only observed sporadically after the known planet orbits
were defined and published \citep{cochran04, endl06}, and 16~Cyg~B has
only been observed with the HJS telescope at a frequency of at most once
per month.  The companion limits are shown in Figures~4-6; planets with
masses above the solid line can be ruled out by the data with 99\%
confidence.  Not surprisingly, the tightest limits were obtained for
HD~3651 (Figure~4), which had a total of 195 measurements, including 29
independent HET visits.  For periods less than about 1~year, we can
exclude planets with M~sin~$i$~\gtsimeq~2 Neptune masses.  Similar results
were obtained for 16~Cyg~B (N=161), where the limits approach a Neptune
mass (Figure~6).  Since the detection limits generally improve with the
addition of more data and with higher-quality data, we can define a
quantity to measure the goodness of the limits.  A simple choice would be
$N/\bar{\sigma}$, where $N$ is the total number of observations, and
$\bar{\sigma}$ is the mean uncertainty of the radial-velocity
measurements.  The values of $N$ and $\bar{\sigma}$ are given in Table~2.

In the HD~45350 system, the results of the dynamical simulations
complement those of the detection limit determinations. Very tight limits
are obtained in close orbits ($a\ltsimeq$0.2~AU).  In this region, test
particles were stable (Fig.~2) and our observations can exclude planets
with M~sin~\textit{i} between about 1 and 4 Neptune masses.  Similar
results were obtained for the 16~Cyg~B system, in which test particles
remained stable inward of $a\sim$~0.3~AU.  In that region, planets of 1-3
Neptune masses can be excluded by our limits determinations (Fig.~6). In
most of the limits determinations, there are multiple ``blind spots''
evident where the periodogram method failed to significantly recover the
injected signals.  Typically this occurs at certain trial periods for
which the phase coverage of the observational data is poor, and often at
the 1-month and 1-year windows.

For none of HD~37605 (Fig.~4), HD~80606 (Fig.~5), or HD~89744 (Fig.~6)
could additional companions be ruled out below about 0.7 \Mjup, and for
most orbital periods tested, the limits were substantially worse.  One
possible explanation for this result is that the sampling of the
observations was poorly distributed in phase for many of the injected test
signals, making significant recovery by the periodogram method difficult.  
This is evidenced by the ``jagged'' regions in the plots.  Also, the
intrinsic scatter for those three systems was too large to permit tight
limits determination.  This is certainly reasonable for the F7 star
HD~89744.  The three systems with the best limits (HD~3651, HD~45350, and
16~Cyg~B) also had the lowest rms scatter about their orbital solutions
(mean=$8.9\pm1.4$ \ms;  Table~1).  In contrast, the mean rms for the
remaining three systems was $13.7\pm0.6$ \ms.  Additional factors such as
a paucity of data (HD~37605) and short time baselines (HD~80606, HD~89744)
made the determination of useful companion limits challenging for some of
the planetary systems in this study.

\section{Summary}

We have shown that for a sample of six highly eccentric extrasolar
planetary systems, there is no evidence for additional planets.  Test
particle simulations show that there are regions detectable by current
surveys (i.e.~for $a<2$ AU) where additional objects can exist. For
HD~3651 and HD~37605, we find that protected resonances are also present.  
Combining these simulations with detection limits computed using new
high-precision HET data combined with all available published data is
particularly effective for the HD~3651 and HD~45350 systems.  Additional
short-period planets can be ruled out down to a few Neptune masses in the
dynamically stable regions in these systems.

\acknowledgements

This material is based upon work supported by the National Aeronautics and
Space Administration under Grant Nos.~NNG04G141G and NNG05G107G issued
through the Terrestrial Planet Finder Foundation Science program.  We are
grateful to the HET TAC for their generous allocation of telescope time
for this project.  We also would like to thank Barbara McArthur for her
assistance with GaussFit software.  We thank the referee, Greg Laughlin,
for his careful review of this manuscript.  This research has made use of
NASA's Astrophysics Data System (ADS), and the SIMBAD database, operated
at CDS, Strasbourg, France.  The Hobby-Eberly Telescope (HET) is a joint
project of the University of Texas at Austin, the Pennsylvania State
University, Stanford University, Ludwig-Maximilians-Universit{\" a}t
M{\"u}nchen, and Georg-August-Universit{\" a}t G{\" o}ttingen The HET is
named in honor of its principal benefactors, William P.~Hobby and Robert
E.~Eberly.

\clearpage



\clearpage

\begin{deluxetable}{lllllllll}
\tabletypesize{\scriptsize}
\tablecolumns{9}
\tablewidth{0pt}
\tablecaption{Keplerian Orbital Solutions \label{tbl-1}}
\tablehead{
\colhead{Planet} & \colhead{Period } & \colhead{$T_0$ } &
\colhead{$e$} & \colhead{$\omega$} & \colhead{K } & \colhead{M
sin $i$ } & \colhead{$a$ } & \colhead{rms }\\
\colhead{} & \colhead{(days)} & \colhead{(JD-2400000)} & \colhead{} &
\colhead{(degrees)} & \colhead{(\ms)} & \colhead{(\Mjup)} & \colhead{(AU)} 
& \colhead{\ms}
 }
\startdata
HD 3651 b & 62.197$\pm$0.012 & 53932.2$\pm$0.4 & 0.630$\pm$0.046 &
250.7$\pm$6.3 & 15.6$\pm$1.1 & 0.20$\pm$0.01 & 0.280$\pm$0.006 & 7.1 \\
HD 37605 b & 55.027$\pm$0.009 & 52992.8$\pm$0.1 & 0.677$\pm$0.009 &
218.4$\pm$1.7 & 201.5$\pm$3.9 & 2.39$\pm$0.12 & 0.263$\pm$0.006 & 13.0 \\
HD 45350 b & 963.6$\pm$3.4 & 51825.3$\pm$7.1 & 0.778$\pm$0.009 &
343.4$\pm$2.3 & 58.0$\pm$1.7 & 1.79$\pm$0.14 & 1.92$\pm$0.07 & 9.1 \\
HD 80606 b & 111.428$\pm$0.002 & 53421.928$\pm$0.004 & 0.933$\pm$0.001 &
300.4$\pm$0.3 & 470.2$\pm$2.5 & 4.10$\pm$0.12 & 0.460$\pm$0.007 & 13.5 \\
HD 89744 b & 256.78$\pm$0.05 & 53816.1$\pm$0.3 & 0.689$\pm$0.006 &
194.1$\pm$0.6 & 263.2$\pm$3.9 & 7.92$\pm$0.23 & 0.91$\pm$0.01 & 14.4 \\
16 Cyg B b & 799.5$\pm$0.6 & 50539.3$\pm$1.6 & 0.689$\pm$0.011 &
83.4$\pm$2.1 & 51.2$\pm$1.1 & 1.68$\pm$0.07 & 1.68$\pm$0.03 & 10.6 \\
\enddata
\end{deluxetable}

\begin{deluxetable}{lllll}
\tabletypesize{\scriptsize}
\tablecolumns{5}
\tablewidth{0pt}
\tablecaption{Summary of Radial-Velocity Data \label{tbl-2}}
\tablehead{
\colhead{Star} & \colhead{$N$} & \colhead{$\bar{\sigma}$ (\ms)} &
\colhead{$\Delta T$ (days)} & \colhead{Source}
}
\startdata
HD 3651 & 163 & 3.4 &  & \citet{butler06} \\
HD 3651 & 3 & 6.1 &  & HJS\tablenotemark{a} \\
HD 3651 & 29 & 2.1 &  & HET\tablenotemark{b} \\
HD 3651 (total) & 195 & 3.2 & 7083 & \\
\hline
HD 37605 (total) & 43 & 2.9 & 1065 & HET \\
\hline
HD 45350 & 38 & 2.8 &  & \citet{butler06} \\
HD 45350 & 28 & 4.2 &  & HET \\
HD 45350 & 47 & 8.9 &  & HJS \\
HD 45350 (total) & 113 & 5.7 & 2265 & \\
\hline
HD 80606 & 61 & 13.7 &  & \citet{naef01} \\
HD 80606 & 46 & 5.1 &  & \citet{butler06} \\
HD 80606 & 23 & 2.5 &  & HET \\
HD 80606 (total) & 130 & 8.7 & 2893 & \\
\hline
HD 89744 & 50 & 11.2 &  & \citet{butler06} \\
HD 89744 & 33 & 3.2 &  & HET \\
HD 89744 & 6 & 9.4 &  & HJS \\
HD 89744 (total) & 89 & 8.1 & 2687 & \\
\hline
16 Cyg B & 95 & 6.3 &  & \citet{butler06} \\
16 Cyg B & 29 & 19.7 &  & HJS Phase II\tablenotemark{c} \\
16 Cyg B & 37 & 7.4 &  & HJS Phase III \\
16 Cyg B (total) & 161 & 9.0 & 6950 & \\
\enddata
\tablenotetext{a}{McDonald Observatory 2.7~m Harlan J.~Smith Telescope.}
\tablenotetext{b}{McDonald Observatory 9.2~m Hobby-Eberly Telescope.}
\tablenotetext{c}{Phase~II indicates an earlier instrument setup detailed 
in \citet{limitspaper}. Phase~III is the current configuration. }
\end{deluxetable}

\clearpage


\begin{figure}
\plottwo{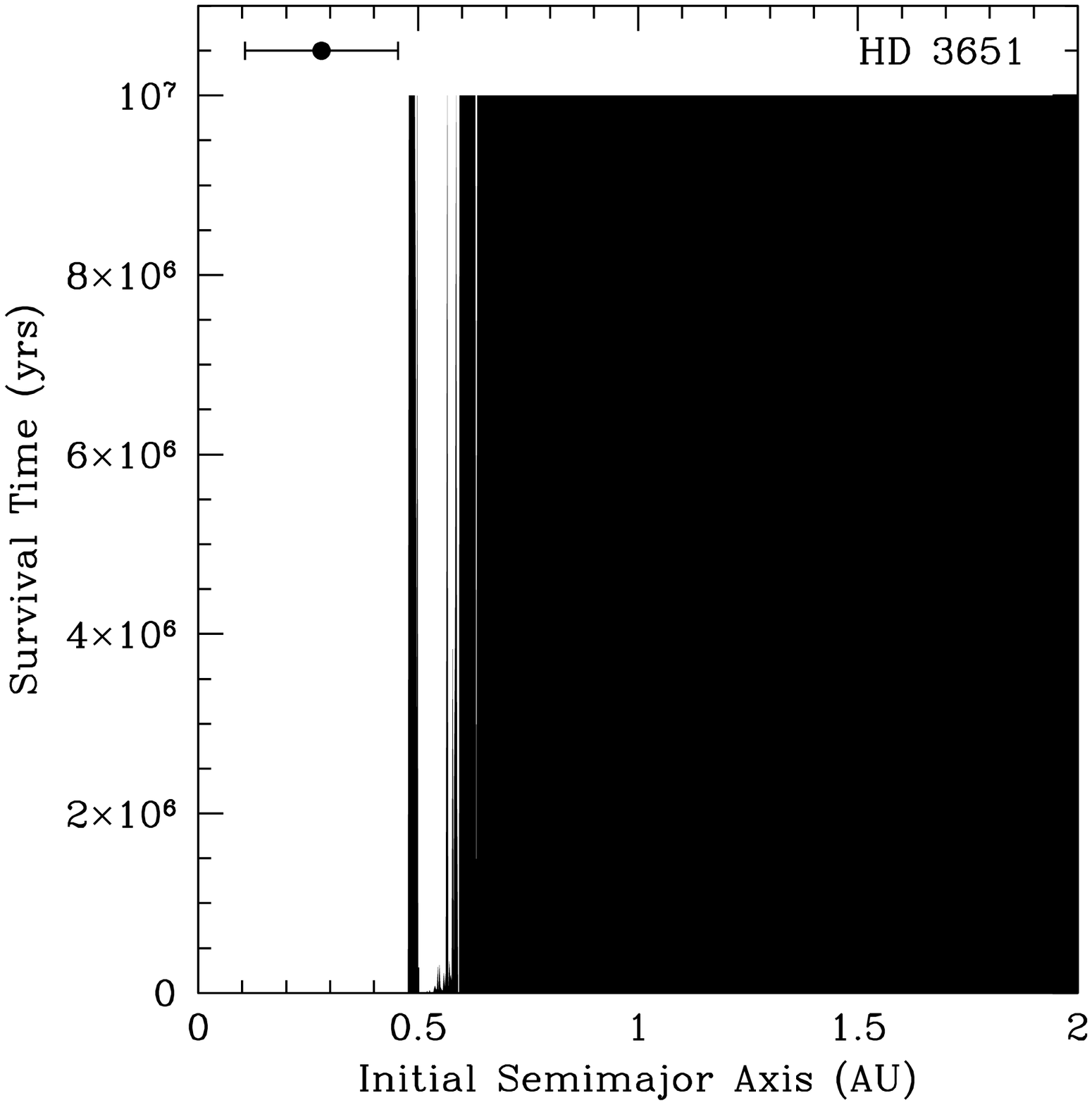}{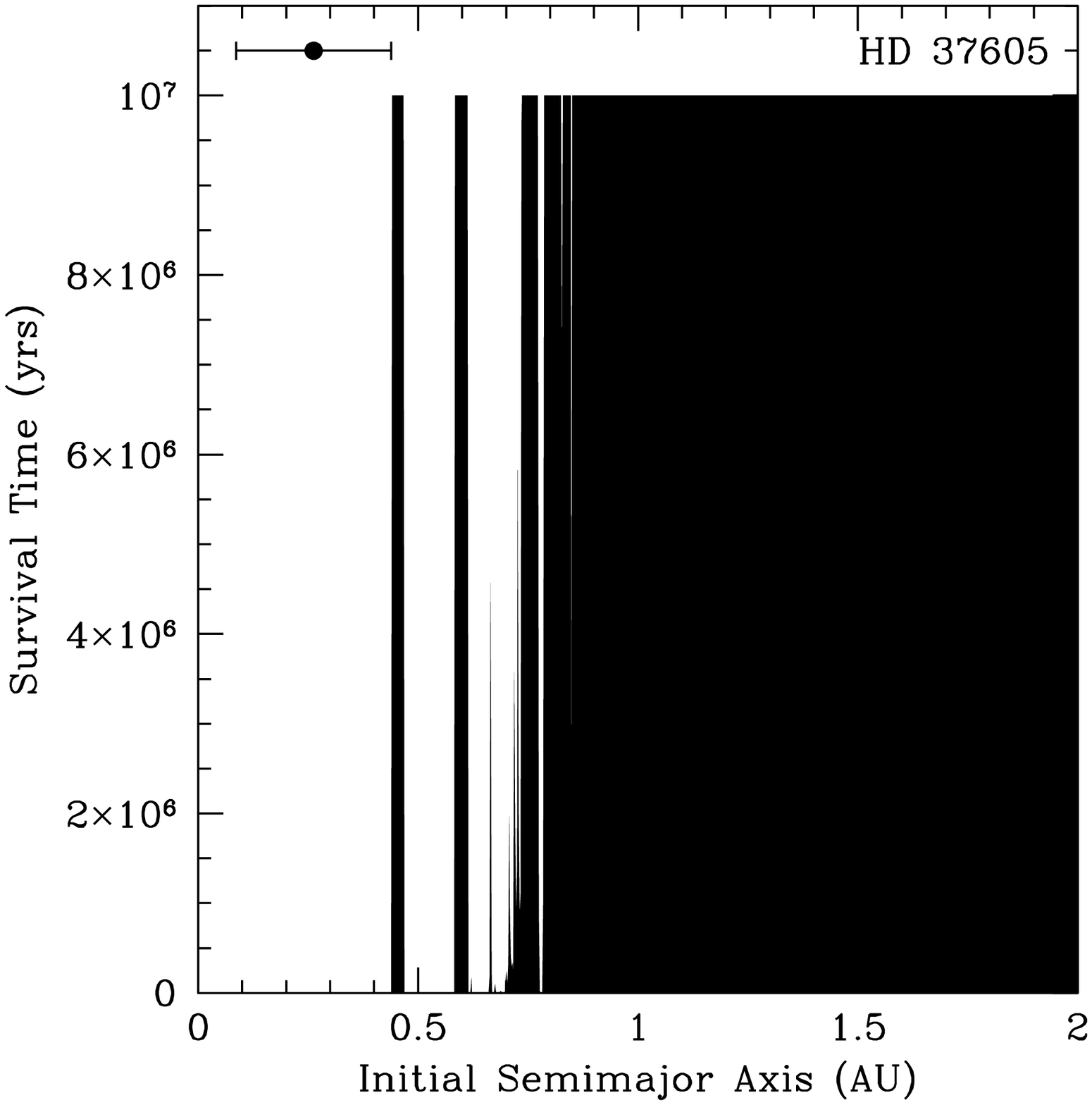}
\caption{Left panel: Survival time as a function of initial semimajor axis
for test particles in the HD~3651 system after $10^7$ yr.  The filled
regions indicate test particles which survived.  The orbital excursion of
HD~3561b is indicated by the horizontal error bars at the top.  Particles
were placed on initially circular orbits with $0.05<a<2.00$~AU.  For all
systems, the known planet removed particles which crossed its orbit. The
dark region near 0.5~AU shows the stable 1:2 mean-motion resonance (MMR).  
Right panel: Same, but for the HD~37605 system. The dark regions near
0.45~AU and 0.6~AU show the stable 1:2 and 1:3 MMRs. }
\end{figure}

\begin{figure}
\plottwo{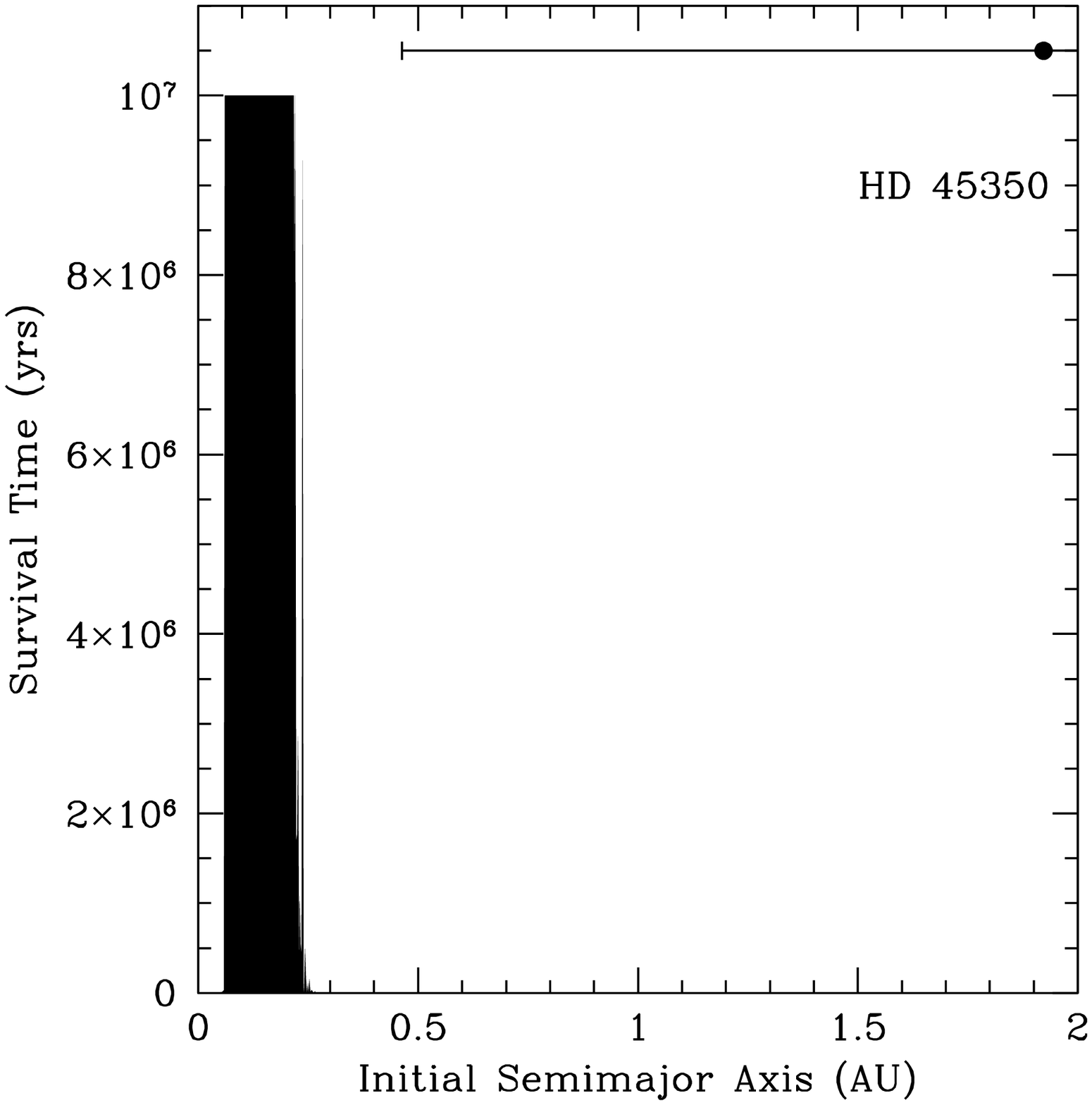}{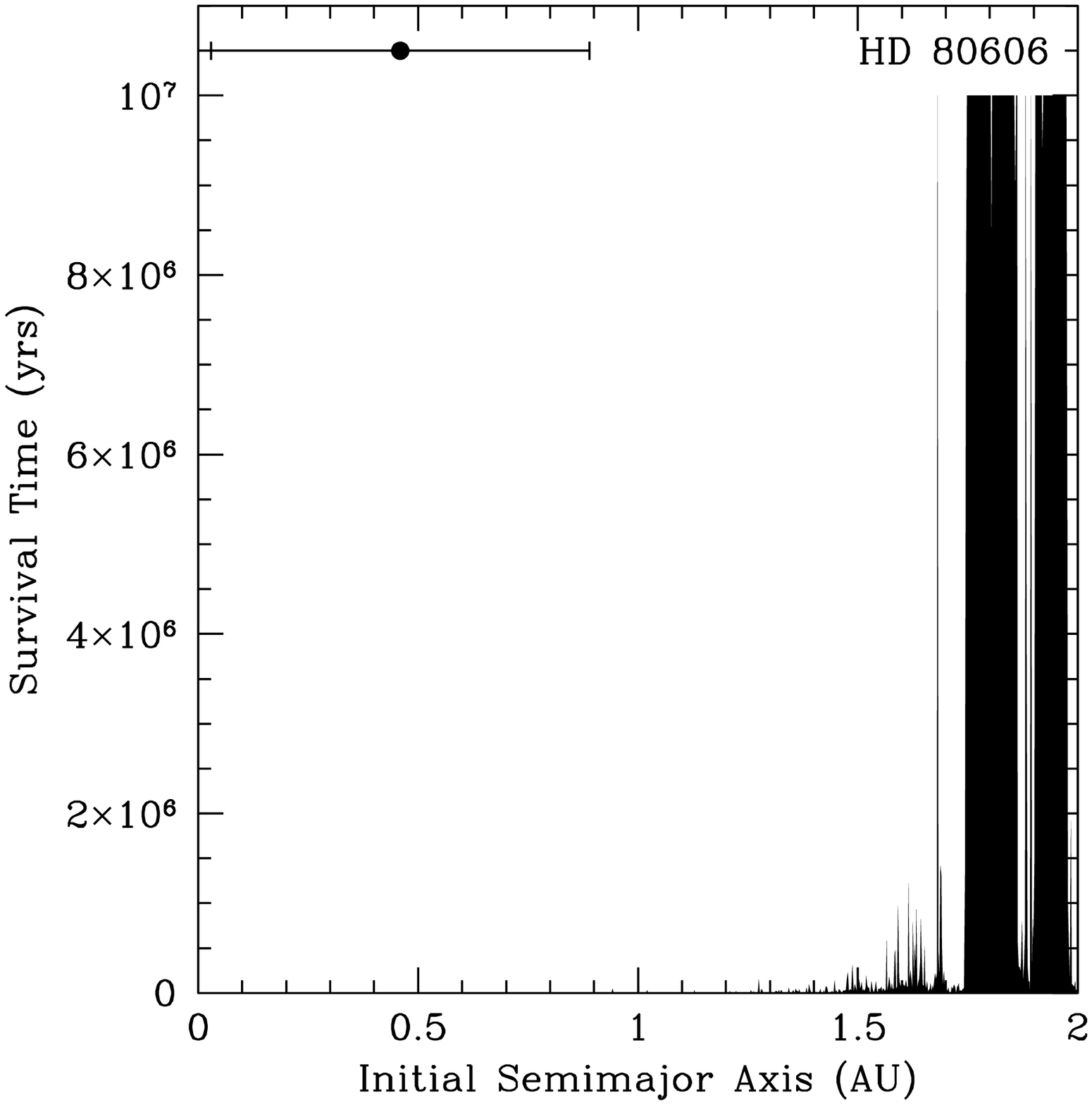}
\caption{Same as Fig.~1, but for the HD~45350 (left) and HD~80606 
(right) systems. }
\end{figure}

\begin{figure}
\plottwo{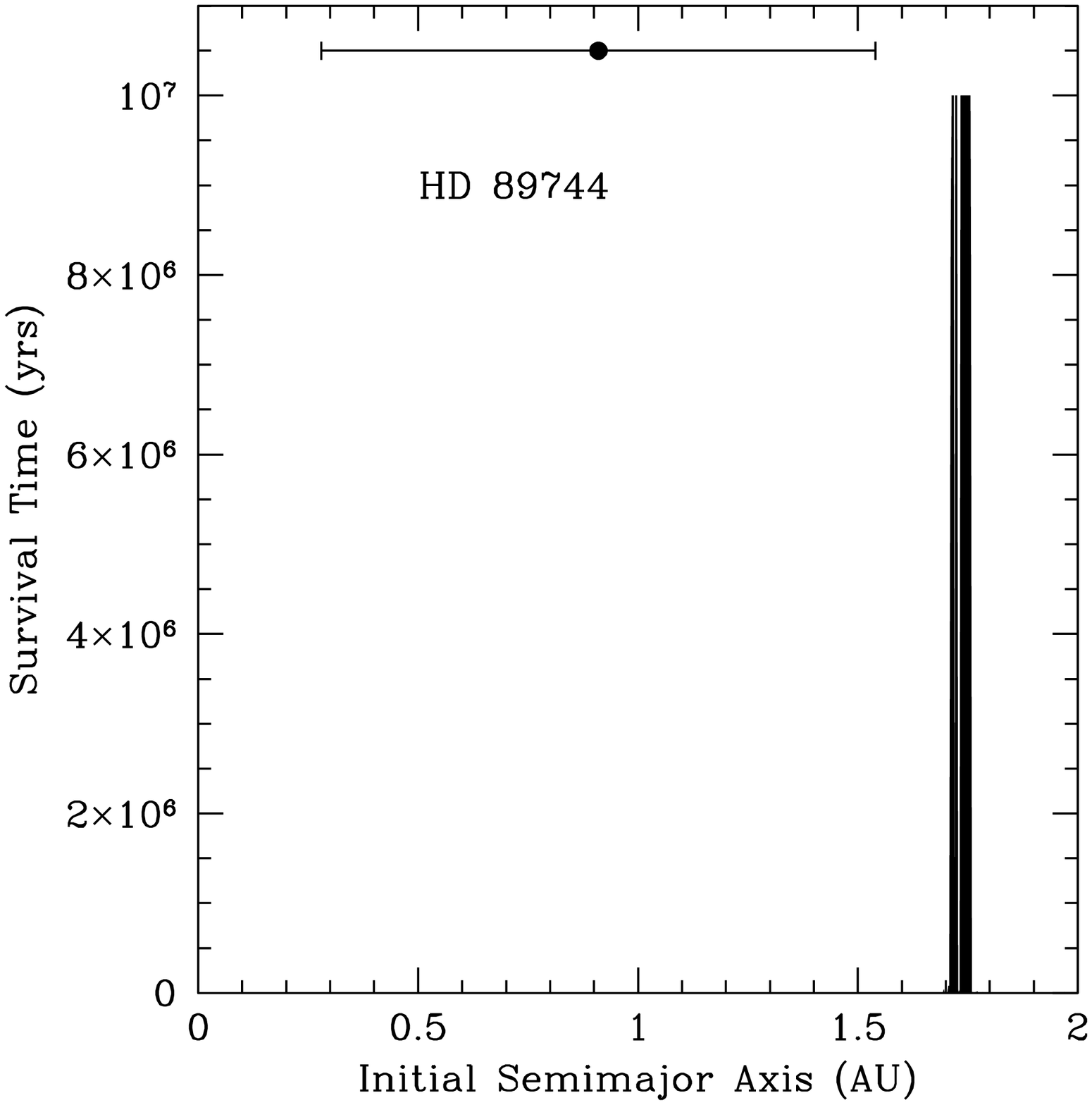}{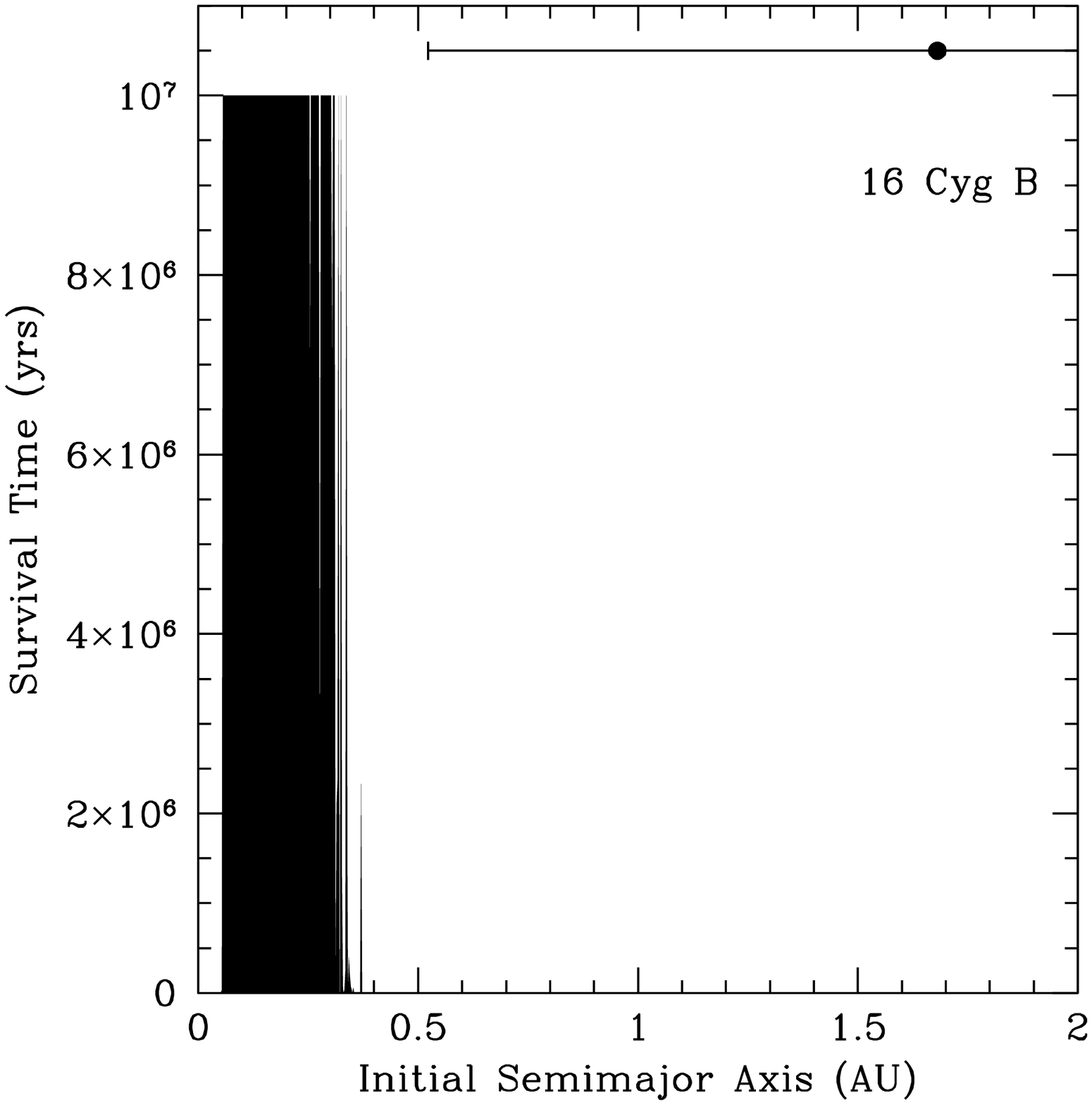}
\caption{Same as Fig.~1, but for the HD~89744 (left) and 16~Cyg~B (right)
systems. }
\end{figure}

\begin{figure}
\plottwo{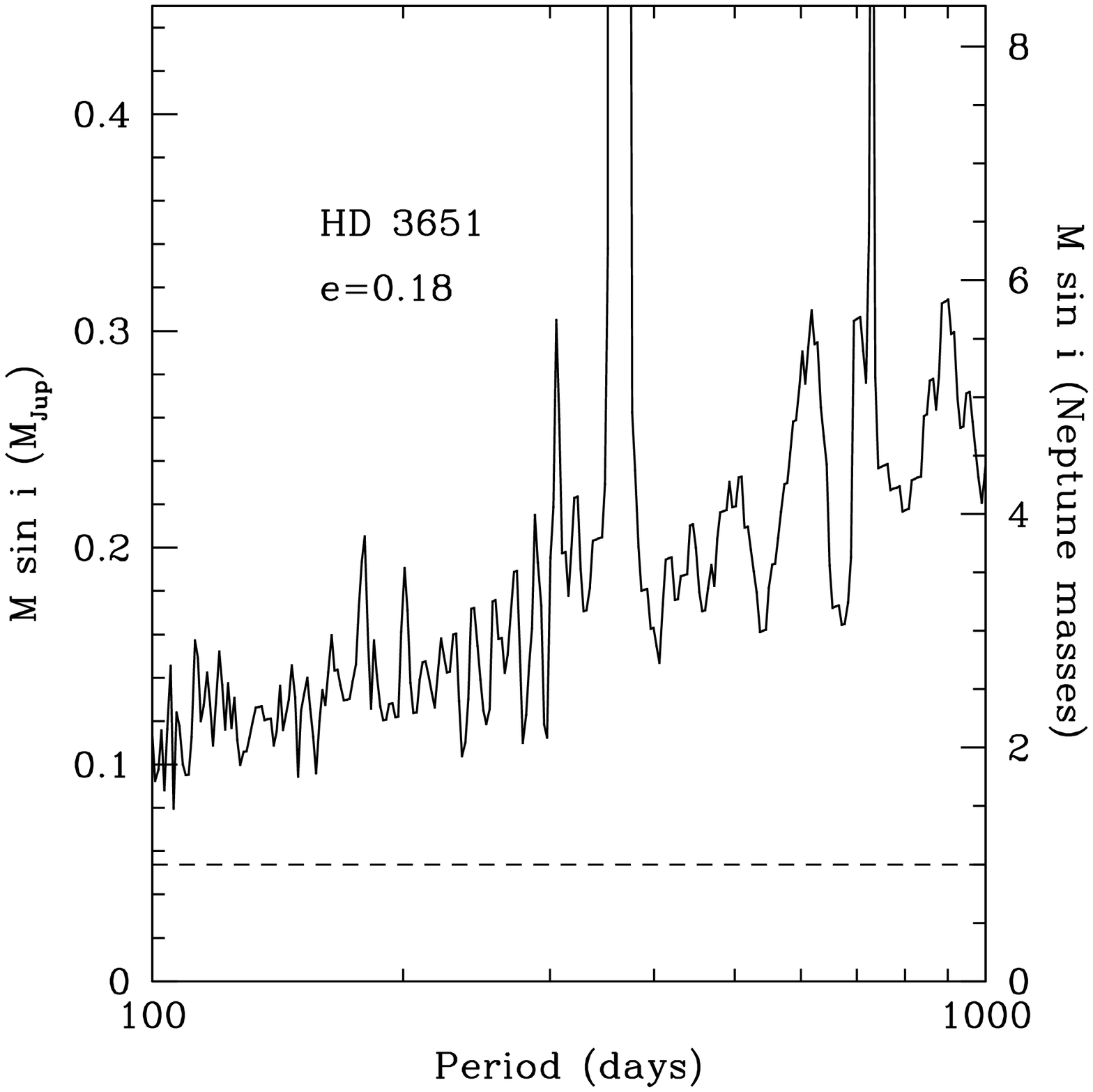}{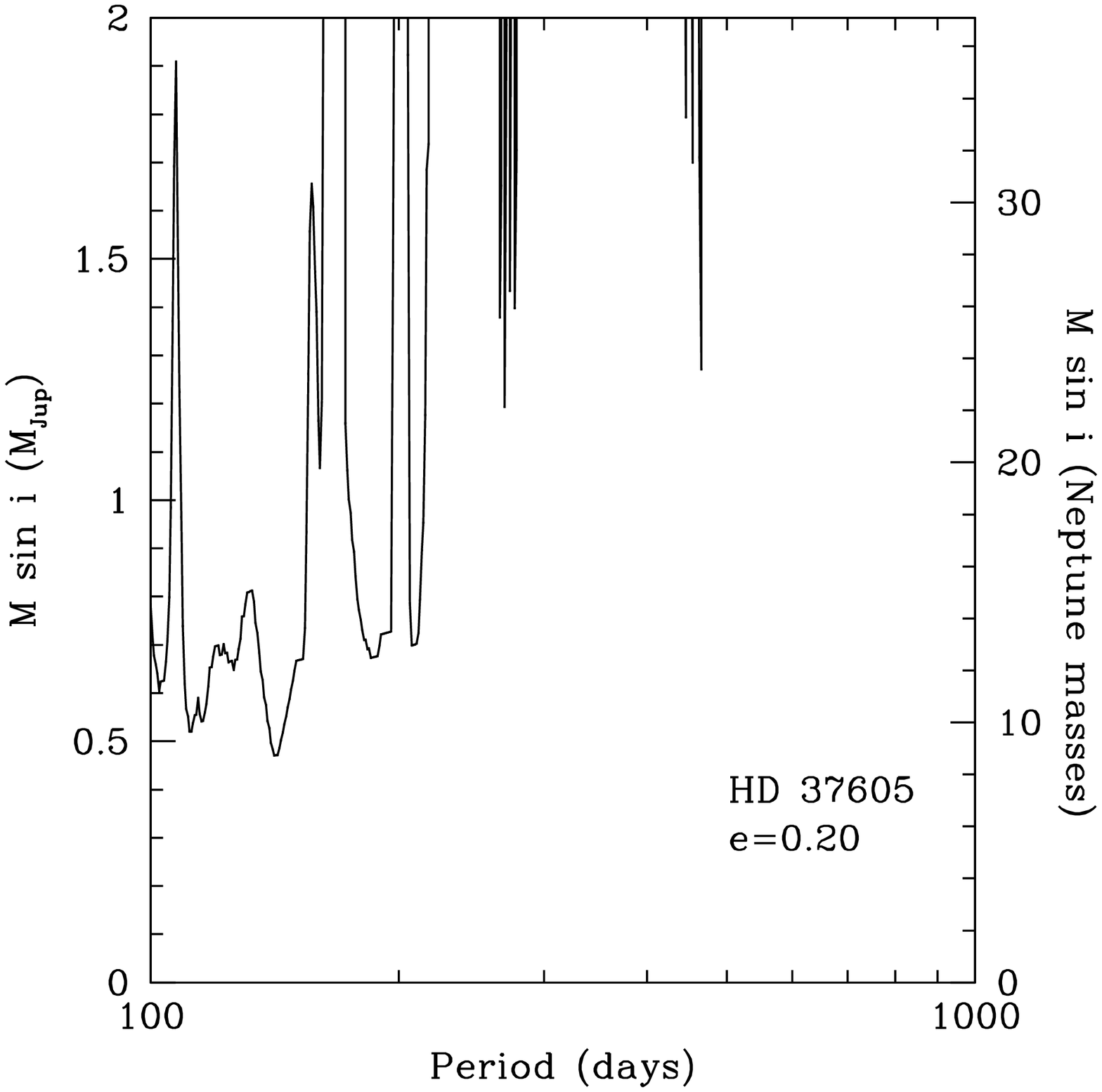}
\caption{Left panel: Detection limits for additional planets in orbits
with $e=0.18$ in the HD~3651 system.  Planets in the parameter space above
the plotted points are excluded at the 99\% confidence level.  The
horizontal dashed line indicates the mass of Neptune.  Right panel: Same,
but for planets with $e=0.20$ in the HD~37605 system. }
\end{figure}

\begin{figure}
\plottwo{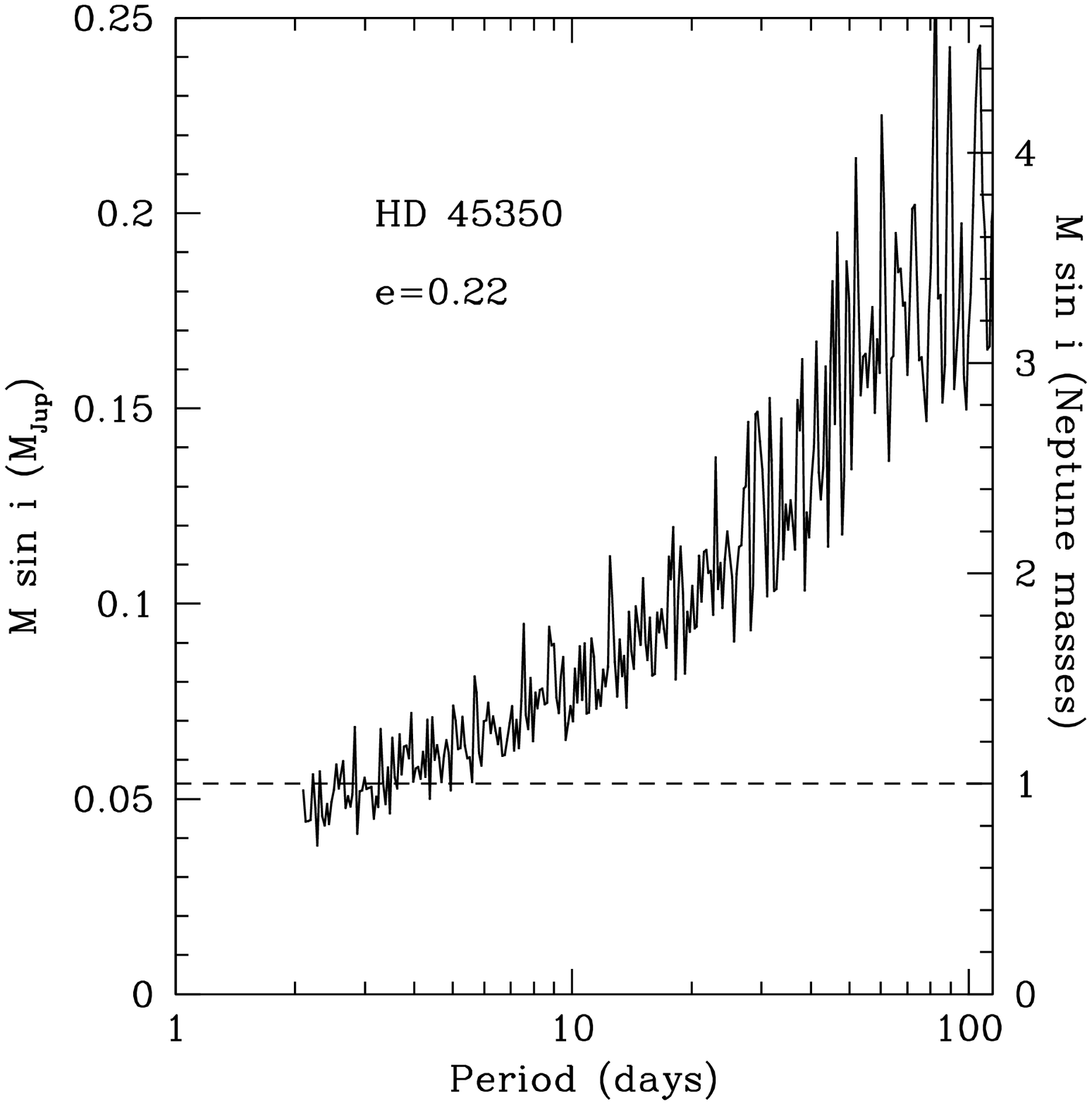}{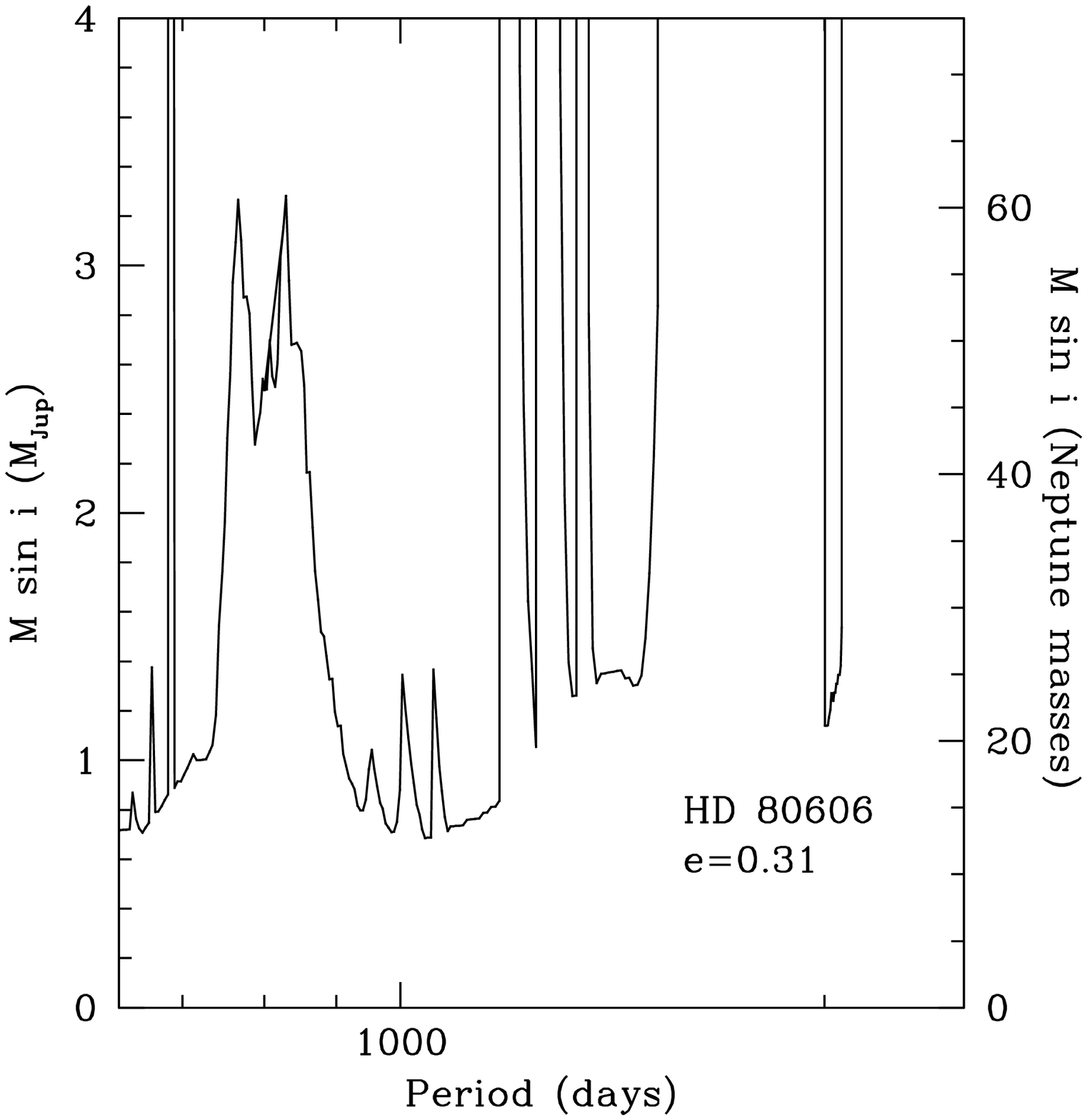}
\caption{Left panel: Detection limits for additional planets with $e=0.22$
in the HD~45350 system.  The horizontal dashed line indicates the mass of
Neptune.  Right panel: Same, but for planets with $e=0.31$ in the HD~80606
system. Planets in the parameter space above the plotted points are
excluded at the 99\% confidence level. }
\end{figure}

\begin{figure}
\plottwo{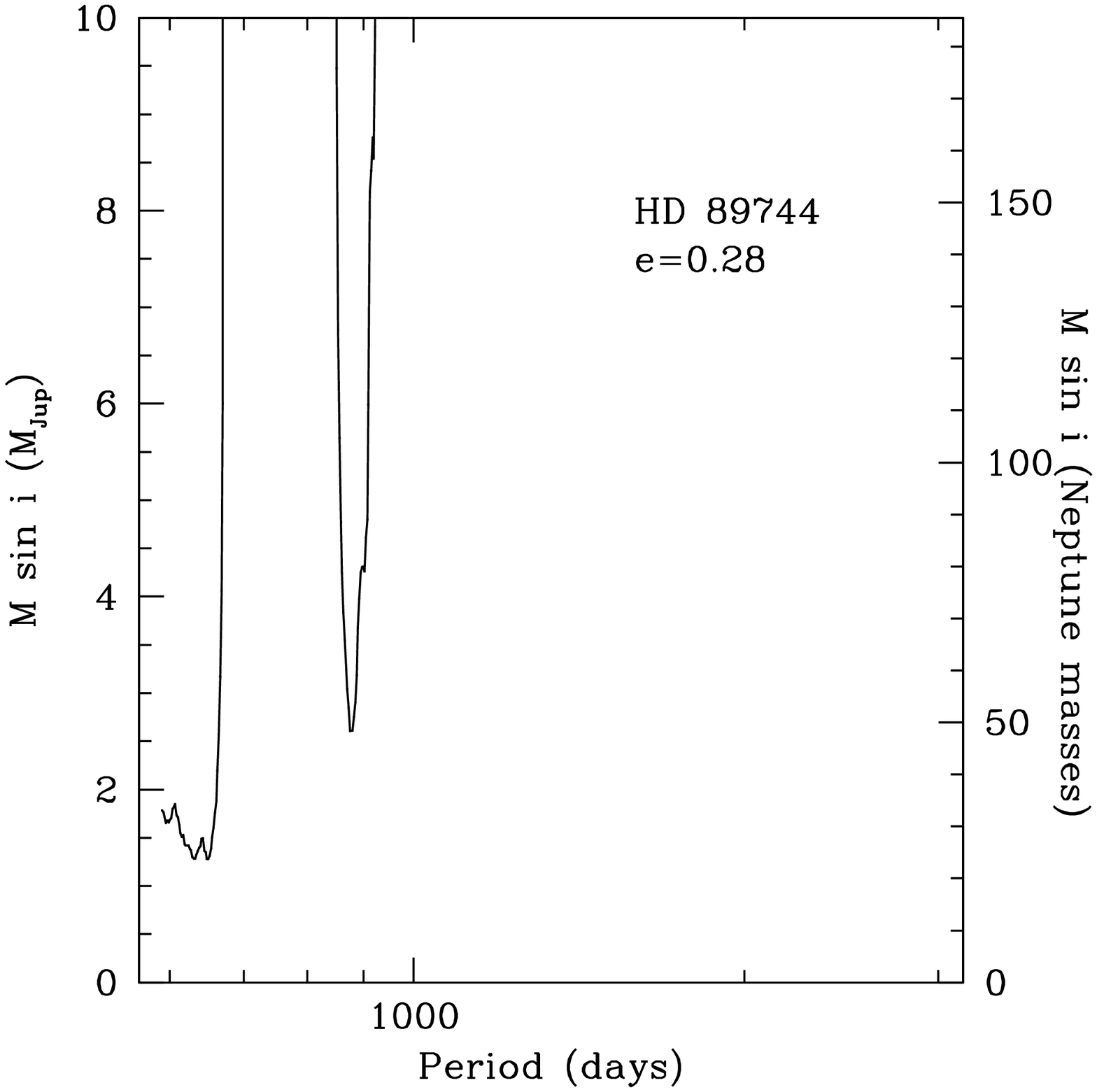}{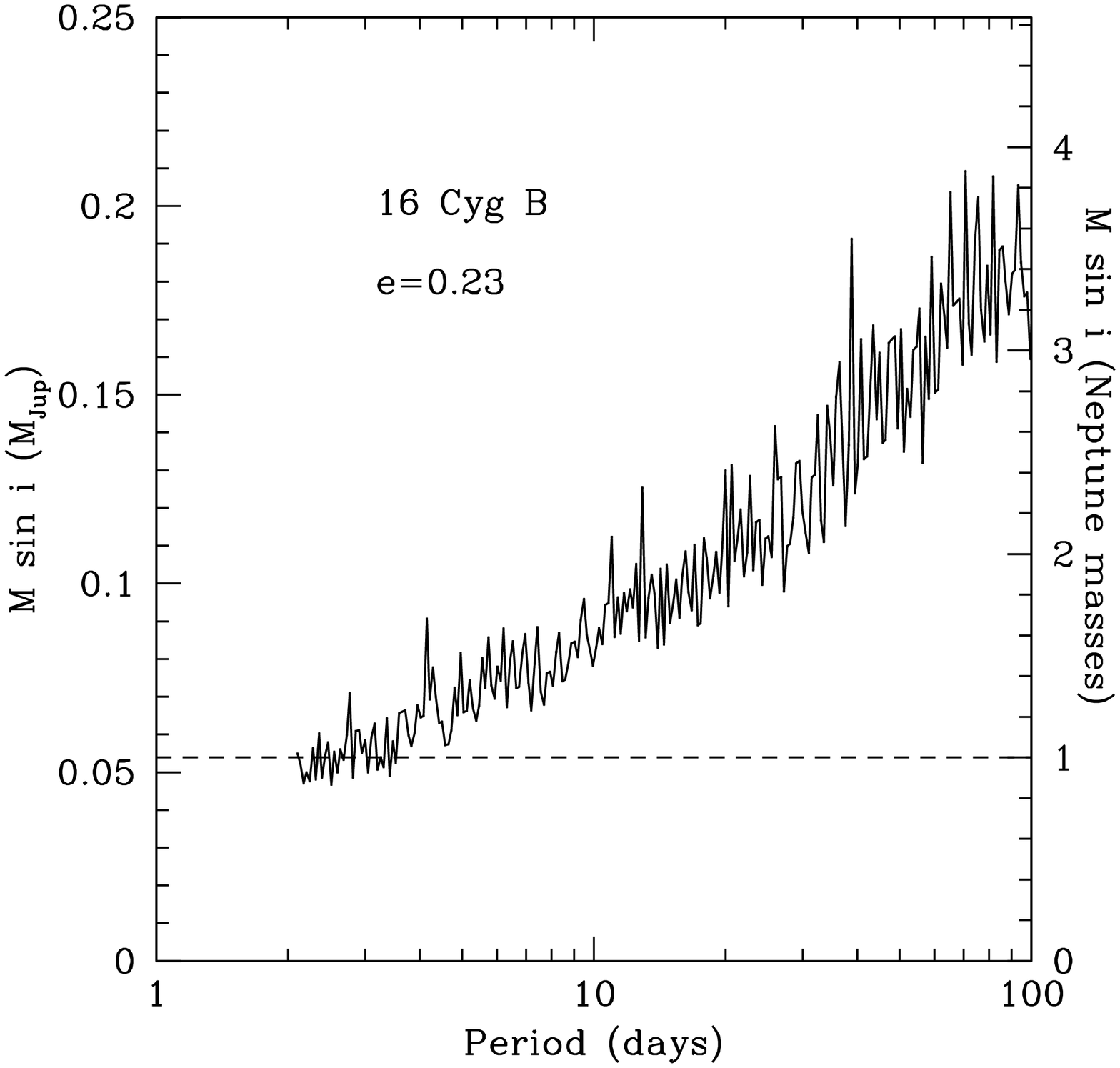}
\caption{Left panel: Detection limits for additional planets with $e=0.28$
in the HD~89744 system.  Right panel: Same, but for planets with $e=0.23$
in the 16~Cyg~B system.  The horizontal dashed line indicates the mass of
Neptune.  Planets in the parameter space above the plotted points are
excluded at the 99\% confidence level. }
\end{figure}

\clearpage

\begin{deluxetable}{lrr}
\tabletypesize{\scriptsize}
\tablecolumns{3}
\tablewidth{0pt}
\tablecaption{HET Radial Velocities for HD 3651}
\label{tbl-3}
\tablehead{
\colhead{JD-2400000} & \colhead{Velocity (\ms)} & \colhead{Uncertainty
(\ms)}}
\startdata
53581.87326  &    -19.1  &    2.9  \\
53581.87586  &    -19.4  &    2.7  \\
53581.87846  &    -20.7  &    2.7  \\
53600.79669  &    -11.5  &    2.4  \\
53600.79860  &    -15.5  &    3.0  \\
53600.80050  &    -22.8  &    2.9  \\
53604.79166  &    -15.8  &    1.9  \\
53604.79356  &    -18.8  &    2.1  \\
53604.79548  &    -21.3  &    2.1  \\
53606.78169  &    -19.3  &    1.8  \\
53606.78360  &    -14.8  &    2.1  \\
53606.78551  &    -24.0  &    1.8  \\
53608.77236  &    -18.8  &    1.9  \\
53608.77426  &    -18.0  &    1.9  \\
53608.77617  &    -18.8  &    1.8  \\
53615.96280  &    -28.0  &    2.6  \\
53615.96471  &    -31.9  &    2.4  \\
53615.96662  &    -37.8  &    2.5  \\
53628.74050  &     -6.8  &    2.2  \\
53628.74240  &    -14.5  &    2.4  \\
53628.74431  &     -5.5  &    2.2  \\
53669.61012  &    -18.2  &    2.1  \\
53669.61203  &    -19.2  &    2.2  \\
53669.61394  &    -17.7  &    2.4  \\
53678.78954  &    -10.6  &    2.4  \\
53678.79141  &     -8.6  &    2.3  \\
53678.79332  &     -2.3  &    2.1  \\
53682.78423  &    -15.4  &    2.2  \\
53682.78609  &    -15.0  &    2.3  \\
53682.78801  &    -11.9  &    2.3  \\
53687.77684  &     11.3  &    2.2  \\
53687.77875  &      8.7  &    2.2  \\
53687.78066  &     15.9  &    2.2  \\
53691.75967  &      9.6  &    2.2  \\
53691.76158  &     20.3  &    2.1  \\
53691.76349  &     15.9  &    2.0  \\
53696.75837  &     16.1  &    1.8  \\
53696.76028  &     18.6  &    1.8  \\
53696.76220  &     20.0  &    2.0  \\
53694.75275  &     18.0  &    1.9  \\
53694.75466  &     15.1  &    2.0  \\
53694.75656  &     17.8  &    2.0  \\
53955.83401  &     -0.5  &    1.9  \\
53955.83593  &     -1.2  &    2.0  \\
53955.83785  &      1.3  &    1.9  \\
53956.82850  &      0.4  &    2.0  \\
53956.83046  &     -1.0  &    2.0  \\
53956.83236  &     -5.4  &    2.2  \\
53957.82201  &     -2.1  &    2.0  \\
53957.82392  &     -1.3  &    2.0  \\
53957.82583  &     -3.6  &    2.0  \\
53973.80721  &      9.8  &    7.3  \\
53973.81020  &      3.5  &    2.3  \\
53973.81200  &     -3.5  &    2.0  \\
53976.78393  &    -10.4  &    2.4  \\
53976.78586  &     -5.4  &    2.1  \\
53976.78778  &     -6.7  &    2.3  \\
53978.97197  &     -3.8  &    2.6  \\
53985.95886  &     -9.0  &    2.3  \\
53985.96079  &      4.3  &    3.3  \\
53987.95335  &     -8.3  &    2.2  \\
53987.95527  &     -8.0  &    2.2  \\
53987.95719  &    -12.0  &    2.3  \\
53989.73817  &    -13.2  &    2.2  \\
53989.74009  &    -13.2  &    2.1  \\
53989.74203  &    -18.6  &    2.1  \\
54003.70719  &      2.0  &    2.2  \\
54003.70915  &      4.7  &    2.4  \\
54005.68297  &      7.0  &    2.5  \\
54005.68488  &     11.1  &    2.0  \\
54005.68690  &     10.2  &    2.1  \\
54056.77919  &     -7.5  &    2.2  \\
54056.78110  &    -11.5  &    2.1  \\
54056.78302  &     -9.6  &    2.3  \\
54062.55119  &     20.1  &    1.8  \\
54062.55312  &     21.9  &    2.0  \\
54062.55505  &     20.9  &    2.0  \\
54064.54710  &     12.8  &    2.0  \\
54064.54902  &     16.7  &    2.1  \\
54064.55094  &     16.6  &    2.1  \\
54130.55316  &     19.1  &    2.4  \\
54130.55508  &     16.9  &    2.5  \\
54130.55701  &     17.6  &    2.5  \\
\enddata
\end{deluxetable}

\begin{deluxetable}{lrr}
\tabletypesize{\scriptsize}
\tablecolumns{3}
\tablewidth{0pt} 
\tablecaption{HET Radial Velocities for HD 37605}
\label{tbl-4}
\tablehead{
\colhead{JD-2400000} & \colhead{Velocity (\ms)} & \colhead{Uncertainty
(\ms)}}
\startdata
53002.67151  &    487.6  &    3.8  \\
53003.68525  &    495.5  &    3.0  \\
53006.66205  &    496.2  &    3.0  \\
53008.66407  &    501.3  &    2.9  \\
53010.80477  &    499.8  &    2.9  \\
53013.79399  &    482.1  &    2.6  \\
53042.72797  &    269.7  &    2.8  \\
53061.66756  &    489.0  &    2.6  \\
53065.64684  &    479.0  &    2.8  \\
53071.64383  &    463.8  &    2.6  \\
53073.63819  &    460.4  &    2.6  \\
53082.62372  &    422.8  &    2.5  \\
53083.59536  &    422.2  &    2.8  \\
53088.59378  &    418.6  &    4.0  \\
53089.59576  &    379.1  &    2.2  \\
53092.59799  &    343.7  &    2.5  \\
53094.58658  &    323.2  &    2.4  \\
53095.58642  &    302.1  &    2.4  \\
53096.58744  &    302.1  &    3.2  \\
53098.57625  &    193.8  &    2.7  \\
53264.95137  &    164.9  &    3.0  \\
53265.94744  &    112.9  &    3.0  \\
53266.94598  &    113.2  &    3.7  \\
53266.95948  &     74.6  &    3.6  \\
53266.97396  &    119.2  &    8.0  \\
53283.92241  &    471.6  &    2.7  \\
53318.81927  &    213.3  &    3.0  \\
53335.92181  &    496.9  &    2.6  \\
53338.90602  &    493.9  &    2.6  \\
53377.81941  &    109.1  &    2.7  \\
53378.81189  &    214.6  &    2.7  \\
53379.80225  &    338.3  &    2.6  \\
53381.64429  &    436.1  &    2.7  \\
53384.64654  &    482.9  &    2.8  \\
53724.85584  &    468.2  &    2.6  \\
53731.69723  &    435.4  &    2.7  \\
53738.67472  &    404.3  &    2.6  \\
53743.81020  &    400.5  &    2.6  \\
53748.64724  &    348.4  &    2.7  \\
54039.85015  &    272.5  &    3.1  \\
54054.96457  &    437.4  &    2.7  \\
54055.95279  &    422.0  &    2.9  \\
54067.76282  &    376.4  &    2.6  \\
\enddata
\end{deluxetable}

\begin{deluxetable}{lrr}
\tabletypesize{\scriptsize}
\tablecolumns{3}
\tablewidth{0pt}
\tablecaption{HET Radial Velocities for HD 80606 }
\label{tbl-5}
\tablehead{
\colhead{JD-2400000} & \colhead{Velocity (\ms)} & \colhead{Uncertainty
(\ms)}}
\startdata
53346.88103  &    -20.8  &    3.0  \\
53358.02089  &    -49.5  &    2.7  \\
53359.82400  &    -60.4  &    3.0  \\
53361.02985  &    -64.7  &    2.5  \\
53365.03079  &    -77.4  &    2.4  \\
53373.98282  &    -88.4  &    3.0  \\
53377.80112  &   -105.5  &    2.4  \\
53379.75230  &   -109.3  &    2.7  \\
53389.74170  &   -115.3  &    2.5  \\
53391.74400  &   -129.4  &    2.4  \\
53395.72763  &   -146.4  &    2.3  \\
53399.72518  &   -158.4  &    2.5  \\
53401.72497  &   -174.7  &    2.7  \\
53414.67819  &   -219.8  &    3.0  \\
53421.85529  &    261.0  &    2.2  \\
53423.86650  &    322.1  &    2.0  \\
53424.85231  &    245.9  &    2.1  \\
53432.87120  &     87.5  &    1.9  \\
53433.60628  &     70.0  &    2.1  \\
53446.79322  &      4.5  &    1.9  \\
54161.85400  &   -109.5  &    2.8  \\
54166.83797  &   -119.3  &    2.4  \\
54186.76189  &   -184.2  &    2.3  \\
\enddata
\end{deluxetable}

\begin{deluxetable}{lrr}
\tabletypesize{\scriptsize}
\tablecolumns{3}
\tablewidth{0pt}
\tablecaption{HET Radial Velocities for HD 89744 }
\label{tbl-6}
\tablehead{
\colhead{JD-2400000} & \colhead{Velocity (\ms)} & \colhead{Uncertainty
(\ms)}}
\startdata
53709.89685  &   -184.5  &    2.3  \\
53723.85188  &   -238.6  &    2.2  \\
53723.85367  &   -238.2  &    2.5  \\
53723.85546  &   -227.7  &    2.3  \\
53727.84394  &   -238.9  &    2.5  \\
53727.84573  &   -244.9  &    2.4  \\
53727.84752  &   -242.9  &    2.6  \\
53736.81887  &   -257.6  &    2.5  \\
53736.82100  &   -248.2  &    2.9  \\
53736.82315  &   -253.4  &    2.4  \\
53738.03261  &   -246.7  &    2.8  \\
53738.03441  &   -243.3  &    2.4  \\
53738.03620  &   -236.0  &    2.5  \\
53738.80860  &   -240.5  &    2.6  \\
53738.81040  &   -258.9  &    2.4  \\
53738.81219  &   -249.3  &    2.5  \\
53734.81795  &   -242.8  &    2.6  \\
53734.81973  &   -243.9  &    2.8  \\
53734.82152  &   -248.5  &    2.4  \\
53742.79119  &   -252.0  &    2.8  \\
53742.79299  &   -257.2  &    2.8  \\
53742.79479  &   -239.7  &    2.8  \\
53751.78199  &   -257.4  &    2.9  \\
53751.78378  &   -263.1  &    2.5  \\
53751.78558  &   -268.0  &    2.3  \\
53753.78155  &   -273.1  &    2.5  \\
53753.78381  &   -278.7  &    2.5  \\
53753.78607  &   -266.4  &    2.4  \\
53755.76038  &   -286.6  &    2.3  \\
53755.76218  &   -266.5  &    2.6  \\
53755.76397  &   -274.9  &    2.7  \\
53746.81506  &   -257.1  &    1.9  \\
53746.81778  &   -250.9  &    2.1  \\
53746.82051  &   -245.2  &    2.3  \\
53757.77002  &   -277.6  &    2.4  \\
53757.77181  &   -280.3  &    2.4  \\
53757.77360  &   -288.7  &    2.2  \\
53797.64609  &   -439.8  &    3.1  \\
53797.64834  &   -462.6  &    2.8  \\
53797.65059  &   -452.5  &    2.9  \\
53809.62428  &   -658.6  &    2.4  \\
53809.62700  &   -658.8  &    2.5  \\
53809.62972  &   -659.2  &    2.3  \\
53837.76359  &   -304.3  &    3.0  \\
53837.76670  &   -324.0  &    2.9  \\
53837.78731  &   -308.6  &    2.7  \\
53837.79077  &   -285.2  &    2.6  \\
53866.69987  &   -215.9  &    1.7  \\
53866.70329  &   -228.3  &    1.7  \\
53866.70670  &   -220.4  &    1.8  \\
53868.68349  &   -251.6  &    3.8  \\
53868.68562  &   -208.6  &    2.9  \\
53868.68777  &   -247.4  &    9.7  \\
53875.66956  &   -215.7  &    1.6  \\
53883.65565  &   -213.8  &    1.8  \\
53883.65837  &   -209.2  &    1.7  \\
53883.66109  &   -200.4  &    1.7  \\
53890.63776  &   -203.4  &    1.7  \\
53890.63954  &   -202.6  &    1.9  \\
53890.64134  &   -203.2  &    1.9  \\
53893.62959  &   -193.8  &    2.0  \\
53893.63139  &   -189.3  &    1.9  \\
53893.63318  &   -189.7  &    1.8  \\
54047.94811  &   -375.2  &    4.8  \\
54047.94991  &   -353.2  &    4.5  \\
54047.95172  &   -362.6  &    4.4  \\
54050.96248  &   -415.0  &    2.6  \\
54050.96453  &   -423.0  &    2.5  \\
54050.96657  &   -420.1  &    2.4  \\
54052.96488  &   -426.8  &    2.3  \\
54052.96762  &   -437.1  &    2.5  \\
54052.97035  &   -447.6  &    2.5  \\
54056.94606  &   -468.0  &    3.0  \\
54056.94786  &   -466.4  &    2.6  \\
54056.94964  &   -479.4  &    2.8  \\
54063.92981  &   -599.1  &    2.1  \\
54063.93166  &   -594.8  &    2.3  \\
54063.93348  &   -592.3  &    2.4  \\
54073.91213  &   -685.8  &    2.8  \\
54073.91476  &   -688.7  &    2.9  \\
54073.91739  &   -704.4  &    2.7  \\
54122.01039  &   -220.8  &    2.5  \\
54122.01243  &   -219.1  &    2.6  \\
54122.01447  &   -218.4  &    2.8  \\
54129.74214  &   -215.7  &    2.6  \\
54129.74491  &   -224.4  &    3.0  \\
54129.74768  &   -223.7  &    3.1  \\
54160.65850  &   -189.5  &    3.2  \\
54160.66031  &   -181.8  &    2.7  \\
54160.66212  &   -204.8  &    3.2  \\
54163.66458  &   -213.9  &    3.1  \\
54163.66643  &   -200.8  &    2.9  \\
54163.66828  &   -208.0  &    3.2  \\
54165.88148  &   -208.5  &    2.7  \\
\enddata
\end{deluxetable}

\end{document}